\documentclass[11pt,a4paper]{article} 
\pdfoutput=1
\usepackage{jcappub}
\usepackage{latexsym,amsmath,amssymb}
\usepackage{graphicx}
\usepackage{slashed}
\usepackage{bm}
\usepackage{epsfig}
\usepackage{dcolumn}

\newcommand{\beq}{\begin{eqnarray}}
\newcommand{\eeq}{\end{eqnarray}}
\newcommand{\be}{\begin{eqnarray}}
\newcommand{\ee}{\end{eqnarray}}
\newcommand{\bea}{\begin{eqnarray}}
\newcommand{\eea}{\end{eqnarray}}

\newcommand{\rar}{\rightarrow}

\def\dd{\text{d} }
\def\d{\partial}
\def\+{\dagger}

\def\<{\langle}
\def\>{\rangle}

\newcommand{\cph}{\varphi}
\newcommand{\cth}{\vartheta}
\newcommand{\cH}{{\cal H}}
\newcommand{\cA}{{\cal A}}

\newcommand{\cL}{{\cal L}}

\title{On inflating magnetic fields, and the backreactions thereof.}

\author{Federico~R.~Urban}

\affiliation{Department of Physics \& Astronomy, University of British Columbia, 6224 Agricultural Road, Vancouver, B.C. V6T 1Z1, Canada}

\date{\today}

\abstract{We investigate in more depth the issue of backreaction in models that attempt at generating cosmological magnetic fields at inflation.  By choosing different, physically motivated, parametrisations, we are able to isolate the heart of the problem, namely the existence, alongside the wanted magnetic field, of its electric counterpart, which turns out quite generally to be stronger and redder.  We were also able to identify a few more interwoven weak spots (the typically very high scale of inflation, the width of the spectrum of modes processed by inflation, the blindness of the amplification mechanism to the energy scale processed), in a way independent on the specifications of the coupling between inflation and electromagnetism.  Despite having stripped down the problem to the core, the obstacles encountered appear insurmountable, thereby posing a challenge to inflation as the incubator of cosmological magnetism.}

\arxivnumber{}
\emailAdd{urban@phas.ubc.ca}

\begin{document}
\maketitle

\section{Introduction}\label{GuessWhat}

The origin of large scale cosmic magnetic fields pertains to the collection of unsolved problems of cosmological particle physics.  Such fields manifest themselves in a variety of environments, from galaxies to clusters to filaments and beyond, with very similar intensity, usually ranging from a few nGauss to hundreds of $\mu$Gauss; they have surprisingly been detected also in regions of the sky, the intergalactic medium, where little to almost no gas is present.  Some useful and comprehensive reviews can be found in~\cite{Kronberg:1993vk,Han:2002ns,Widrow:2002ud,Govoni:2004as,Brandenburg:2004jv,Beck:2008ty,Grasso:2000wj,Dolgov:2003xd,Durrer:2006pc,Kulsrud:2007an,Subramanian:2008tt,Subramanian:2009fu,Kandus:2010nw,Caprini:2011cw,Widrow:2011hs}, where a detailed account of the multifaceted aspects of their detection and of the attempts made toward an organic explanation of their birth and development is given.

While several, more or less contrived, more or less inventive, models relying on purely astrophysical phenomena and resting on conventional field theory can be built with various degrees of success, and while phase transitions constructed around standard or non-standard particle physics theories do give rise to magnetic seeds with very short coherence lengths to be fed to the magnetohydrodynamical plasma, mechanisms operating during an early inflationary epoch of expansion by means of a, direct or indirect, coupling between the field responsible for the de Sitter evolution and photons remain a theorist's favourite, due to relative mathematical simplicity, the obvious appeal of inflation as an anyhow necessary mechanism to generate primordial perturbations, and a natural, automatic solution of the correlation length issue.

Indeed, in surveying the copious variety of models and constructions available, what catches the eye of a reader is the sheer difficulty encountered in trying to stretch high intensity magnetic fields from short to large wavelengths, either through plasma power redistribution mechanisms (inverse cascade) or through violent astrophysical phenomena (ejection of magnetic field lines in the outer space) -- see~\cite{Urban:2009sw} for a possible exception.  Inflation, on the other hand, has it all from the beginning, because any quantum fluctuation will be pulled apart (causally) outside the Hubble scale during de Sitter acceleration, and will comfortably re-enter during a later slowing expansion stage of the evolution of the Universe.

Generating an electromagnetic field during inflation needs however some beyond the standard model input in order to operate, because electromagnetism in four dimensions is conformally invariant, and our Friedmann-Lema\^itre-Robertson-Walker description of the Universe is conformally flat: the combination of the two results in conformal triviality, meaning that the Euler-Lagrange equations of motion for the electromagnetic vector potential can be cast in their Minkowskian form upon a conformal transformation, and would never notice the presence of the non-trivial, time-dependent background.  The only exception to this is the quantum conformal anomaly~\cite{Dolgov:1993vg}, which unfortunately does not appear to be viable.

In spite of all these encouraging news, inflation as a locus for the dawn of magnetic fields suffers from one major drawback, that of backreaction.  What typically happens is that, although initially negligible, the effects of the growing electromagnetic waves onto the background evolution turn out to be so important that de Sitter expansion could not be supported for a period of time long enough to address successfully the problems inflation was brought in to solve in the first place.  What one finds is that in a large class of well justified models, either the produced magnetic fields will be strong, but inflation would end abruptly early, or inflation survives the threat but the magnetic outcome is miserably weak to be of any help in reconciling with observations~\cite{Demozzi:2009fu,Kanno:2009ei,Emami:2009vd}~\footnote{Backreactions were first mentioned in an earlier work by Campanelli~\cite{Campanelli:2008kh}.}.

In this work we want to dwell deeper on this apparently insurmountable issue, in an attempt to better understand and isolate the details which ultimately cause the wreckage of this approach; we wish to proceed in a model-independent way by parametrising the growth of the electromagnetic vector potential without relying on the specifications of the background particle theoretical description; in light of the physics thereby unveiled one may hope to recognise possible and potential avenues circumventing the obstacles obstructing the way to inflationary magnetic fields.

The procedure we adopt in this paper is as follows.  First, in Sec.~\ref{EMevol} we will describe the type of theories we deal with, and swiftly skim over the arguments which show the rise of backreaction; from there we move on to the different parametrisations we study, corresponding to different underlying physics, in Sec.~\ref{cAevol}, and describe the background dynamics in Sec.~\ref{INFevol}.  With all the ingredients at hand we show how much magnetic field can possibly be generated in a variety of situations in Sec.~\ref{results}, detailing on different perspectives in tackling the subject in Secs.~\ref{fixNgetB} and~\ref{fixBgetN}; the discussion of these results is given in Sec.~\ref{some}, while an interesting promising, but ultimately doomed, counterexample is outlined in Sec.~\ref{spiky}.  We review and summarise the findings of this work in Sec.~\ref{wrapup}, alongside proposed future developments.

\section{The development of electric and magnetic fields}\label{EMevol}

Generally inflation and electromagnetism are coupled by means of one of the three interaction terms
\be\label{interaction}
\cL_{\cph\gamma\gamma} \equiv -\frac14 I(\cph)^2 F^2 + \frac14 f(\cph) F \tilde F - \frac12 m(\cph)^2 A^2 \, ,
\ee
where the $F_{\mu\nu} \equiv \partial_\mu A_\nu - \partial_\nu A_\mu$ is the conventional electromagnetic field strength tensor, $\tilde{F}^{\mu\nu} \equiv \frac{1}{2}\eta^{\mu\nu\lambda\kappa}F_{\lambda\kappa}$ (with $\eta^{\alpha\beta\gamma\delta}$ being the totally antisymmetric Levi-Civita tensor with $\eta^{0123}=\sqrt{|-g|}$) is its dual -- we adopt mostly negative metric signature $(+,-,-,-)$.  The non trivial dynamics comes from the time-dependence of the functions $I(\eta)$, $f(\eta)$, and $m(\eta)$ where $\eta$ denotes conformal time and the background inflaton field $\cph$ is hence a function of time only.

The equations of motion for the electromagnetic field, which look like
\be\label{eom}
\nabla_\mu F^\mu_\nu + 2(\d_\mu\ln I) F^\mu_\nu - (\d_\mu f) \tilde{F}^\mu_\nu + m^2 A_\nu = 0 \, ,
\ee
encode the modified dynamics of the system.  In general terms one can then proceed to parametrising the evolution of the $I^2$, $f$, and $m^2$, as a function of conformal time, or the scale factor, and investigate the kind of behaviour the electromagnetic four-potential $A_\mu$ in a given gauge.  The simplest way is to quantise the field (we adhere to the definitions of~\cite{Barrow:2006ch}) and look at the specific evolution, for instance, of the helicity $h=\pm$ Fourier modes $\cA_h$ -- we will always work in Coulomb-transverse gauge for which $(A_\mu)=(0,A_i)$ with $\partial_iA_i=0$.

We instead opt for a different approach, since we are ultimately interested in how strong a generated magnetic field could possibly be, where the only constraint we have is that inflation develops and completes successfully.  We therefore wish to be able to analyse the evolution of the Fourier modes $\cA_h$ detachedly from the details of the form of the interaction term; we seek to bring to the forefront the places where the backreaction comes in, and eviscerate the physics at the core of the problem.  In particular, we want to understand whether there are, in principle, ways for the $\cA_h$ modes to evolve developing a magnetic field at large scales which is appreciable today; we aim at clarifying the relative weights of the electric and magnetic fields, and that of the different modes involved; we further look at the r\^ole of the magnetic (and electric) power spectrum in determining the final result, including the possibility of explicit cutoffs beyond those traditionally assumed in coping with this issue.

Before moving on to the actual discussion, the key definitions we will employ in the rest of the paper are the magnetic power spectrum
\be\label{deltaB}
\delta_B^2 \equiv \sum_h \frac{k^5 |\cA_h|^2}{4\pi^2 a^4} \, ,
\ee
where $k$ is the wave number of the given Fourier mode $\cA_h$, and $a(\eta)$ is the scale factor of the Universe.  The energy density in electromagnetic field is expressed as
\be\label{energy}
\rho_\gamma = \sum_h \frac{1}{4\pi^2 a^4} \int \frac{\dd k}{k} k^3 \left[ |\cA_h'|^2 + k^2 |\cA_h|^2 \right] \, ,
\ee
where the first term is related to the electric field, and the latter is the magnetic energy density proper.  The vacuum normalisation for $\cA_h$ is
\be\label{vacuum}
\cA_h^{vac} = \frac{1}{\sqrt{2k}} e^{-ik\eta} \, .
\ee

During the quasi de Sitter expansion epoch, conformal time, scale factor, number of e-foldings and the various Hubble parameters are related as
\be\label{defs}
a = a_i e^N \, , \quad \eta = \eta_i e^{-N} \, , \quad N \equiv \int H_I \dd t \, , \quad \cH \equiv \frac{a'}{a} = a H_I = \dot{a} \, ,
\ee
where $t$ is cosmic time, the scale factors $a_i$, $a_e$, and $a_k$ tell us about the beginning and end of inflation, and the moment at which a given mode exits the Hubble scale $a_k = k/H_I$, $H_I$ being the approximately constant Hubble parameter, which we normally compute at the end of inflation.  Also, recall that during inflation conformal time spans the range $[-\infty,0]$, and $\cH = -1/\eta$.  Finally, conformal time derivative is denoted with a prime, the dot being reserved for cosmic time.

Already at this stage we can obtain a very primitive estimate for the suppression of the final magnetic strength in the most basic case.  Indeed, without specifying anything about the coupling in~(\ref{interaction}), and the details of the evolution of $\cA_h$, we find that the most favourable scenario only yields around $10^{-30}$ Gauss on Mpc scales today.  This can be seen as follows.  Imagine the spectrum~(\ref{deltaB}) is some power of $k$ times an amplification factor due to the interaction with the inflaton:
\be\label{deltaBzero}
\delta_B^{2,e} \simeq \text{Amp}^2 H_e^4 \left( \frac{k}{\cH_e} \right)^{4+n} \, ,
\ee
which translates in magnetic and electric energy densities
\be\label{energyzero}
\rho_B^e \simeq \text{Amp}^2 \frac{H_e^4}{4+n} \left[ 1 - \left( \frac{a_i}{a_e} \right)^{4+n} \right] \, , \quad \rho_E^e \simeq \text{Amp}^2 \frac{H_e^4}{2+n} \left[ 1 - \left( \frac{a_i}{a_e} \right)^{2+n} \right] \, ,
\ee
respectively; notice how the electric part is either dominating or equal to the magnetic one.

There are two distinct possibilities, for either $n>-2$ or $n<-2$.  In the first case the ratio between the photons energy density and the background one is $\rho_E^e / \rho_\cph^e \simeq \text{Amp}^2 H_I^2 / M_4^2$, which has to stay below one.  This implies a maximal amplification factor of $\text{Amp} = M_4/H_I$, and, assuming radiation dominated the energy budget of the Universe at $a_e$, a magnetic field strength today
\be\label{deltaBzero0A}
\delta_B^{2,0} \simeq M_4^2 H_I^2 \left( \frac{k}{\cH_e} \right)^{4+n} \simeq \rho_\text{cmb} \left( \frac{k}{\cH_e} \right)^{4+n} \, .
\ee
If we translate this in Gauss units we find approximately
\be\label{deltaBzero0numA}
\delta_B^0 \simeq 10^{-6} \left( \frac{k}{\cH_e} \right)^{2+n/2} \text{Gauss} \, ,
\ee
where the most favourable case is for $n\rar-2$.  The very wide stretch between large scale $k$, say 1/Mpc, and $\cH_e$ then translates in the $10^{-30}$ Gauss upper limit mentioned before.  The $E$ and $B$ spectra in this case are both very blue.

The other possibility is that $n<-2$, which means that the maximal amplification factor is $\text{Amp} = (a_e/a_i)^{2+n} M_4/H_I$; hence
\be\label{deltaBzero0B}
\delta_B^{2,0} \simeq \rho_\text{cmb} \left( \frac{a_e}{a_i} \right)^{2+n} \left( \frac{k}{\cH_e} \right)^{4+n} \lesssim \rho_\text{cmb} \left( \frac{k}{\cH_0} \right)^{2+n} \left( \frac{k}{\cH_e} \right)^2 \, ,
\ee
where in the last step we have reasoned that inflation has to last enough to at least include the scale $1/H_0$.  Again, in numbers this means
\be\label{deltaBzero0numB}
\delta_B^0 \simeq 10^{-6} \left( \frac{k}{\cH_0} \right)^{1+n/2} \left( \frac{k}{\cH_e} \right) \text{Gauss} \, .
\ee
To minimise the suppression factor $k/H_0$ the best choice is once more $n\rar-2$, for which~(\ref{deltaBzero0A}) is recovered, with identical outcome.  It is important to realise that the magnetic spectrum ultimately wants to be blue ($n=-2$) which is why large scale fields are weak; this constraint comes from the rampant electric field, as we will discuss in details in Sec.~\ref{some}.

\subsection{Parametrising the evolution of $\cA$}\label{cAevol}

\paragraph{Model 1.} Complying with the statement of intentions enunciated above, we identify four parametrisations which capture different aspects of the problem.  The first possibility is the most intuitive one, and includes the analysis of~\cite{Demozzi:2009fu,Kanno:2009ei,Emami:2009vd}:
\be\label{cAmod1}
\cA_h = \frac{1}{\sqrt{2k}} \left\{ \left( \frac{a}{a_i} \right)^p \cth_\text{in} + \left( \frac{a_k}{a_i} \right)^p \left( \frac{a}{a_k} \right)^q \cth_\text{out} \right\} e^{-ik\eta} \, ,
\ee
where we introduced the Heaviside functions $\cth_\text{in} \equiv \cth(|k\eta| - 1)$ and $\cth_\text{out} \equiv \cth(1 - |k\eta|)$ to select the modes that are inside or outside the Hubble scale, respectively.  The parameters set $(p,q)$ describe the growth, or decay, of the vector potential in these two cases; this parametrisation is therefore mode-blind, in the sense that all modes are treated equally, except for their time spent on the two sides of the Hubble horizon, which is obviously $k$-dependent.  The $(p,q) = (0,q)$ case is the most common one arising from interaction terms of the first and third type in~(\ref{interaction}), which normally is inefficient for subhorizon modes; here we have extended it to include non-trivial evolution within the Hubble scale as well.  A convenient way to rewrite~(\ref{cAmod1}) is
\be\label{cAmod1N}
\cA_h = \frac{1}{\sqrt{2k}} \left\{ e^{pN} \cth_\text{in} +|k\eta_i|^{p-q} e^{qN}\cth_\text{out} \right\} e^{-ik\eta} \, ,
\ee
where we can further rework the last term as $|k\eta_i|^{p-q} e^{qN} = |k\eta|^{p-q} e^{pN}$.

At the end of inflation the power spectrum~(\ref{deltaB}) and the total electromagnetic energy density~(\ref{energy}) will be given by
\be\label{deltaBmod1}
\delta_B^{2,e} = \frac{H_I^4}{8\pi^2} \left( \frac{k}{\cH_e} \right)^{4 + 2p - 2q} e^{pN_t} \, ,
\ee
where $N_t$ determines the total number of e-foldings of inflation and $\cH_e = a_e H_I$, and
\be\label{energymod1}
\rho_\gamma^e = \frac{H_I^4}{8\pi^2} \left\{ \frac{1}{2+p-q} \left[ e^{2pN_t} - e^{(2q - 4)N_t} \right] + \frac{q^2}{2+2p-2q} \left[ e^{2pN_t} - e^{(2q - 2)N_t} \right] \right\} \, ,
\ee
where the magnetic field energy density is half of the first term in the curly brackets.  Notice the different exponents in the two terms: they will turn out to be crucial in determining what backreacts and when.  The limits of the $k$ integration are the maximal and minimal $k$ accessible by inflation, that is, $k_\text{max} = a_e H_I = \cH_e$ and $k_\text{min} = a_i H_I = \cH_i$.

\paragraph{Model 2.} The second possibility is related to the axial-type coupling $f F \tilde F$, which is typically efficient only for a short window of time around horizon crossing due to the structure of the mode equation, which includes a friction term of the form $k f' \cA_h'$~\cite{Durrer:2010mq}.  We choose to parametrise this model by
\be\label{cAmod2}
\cA_h &=& \frac{1}{\sqrt{2k}} \left\{ \left[ \cth_\text{pre} + \left( \frac{a}{a_\text{pre}} \right)^r W + \left( \frac{a_\text{post}}{a_\text{pre}} \right)^r \cth_\text{post} \right] \Theta \right.\nonumber\\
&&+ \left. \left[ \left( \frac{a}{a_i} \right)^r W + \left( \frac{a_\text{post}}{a_i} \right)^r \cth_\text{post} \right] \bar\Theta \right\} e^{-ik\eta} \, ,
\ee
where the Heaviside and window functions are
\be
& \cth_\text{pre} \equiv \cth(|k\eta| - c_\text{pre}) \, , \quad \cth_\text{post} \equiv \cth(c_\text{post} - |k\eta|) & , \nonumber\\
& W \equiv \bar\cth_\text{pre} \bar\cth_\text{post} \equiv \cth(c_\text{pre} - |k\eta|) \cth(|k\eta| - c_\text{post}) & , \nonumber\\
& \Theta \equiv \cth(a_\text{pre} - a_i) \, , \quad \bar\Theta \equiv \cth(a_i - a_\text{pre}) & .
\ee
We use the parameters $c_\text{pre} > 1$ and $c_\text{post} < 1$ and the relative Heaviside unit steps $\cth_\text{pre}$ and $\cth_\text{post}$ to describe modes which are not close enough to the Hubble scale, or are far enough out, to be not affected by the growth parametrised by $r$, and confined to the window $W$.  The additional $\Theta$ and $\bar\Theta$ are needed in order to ensure that the amplification window does not begin before $a_i$.  We have further defined $a_\text{pre} \equiv k / c_\text{pre} H_I$ and $a_\text{post} \equiv c_\text{post} k / H_I$.  In what follows, for simplicity, we will choose $c_\text{pre} = c = 1/c_\text{post}$; notice that we demand that at least once the full window $a_\text{pre}$ to $a_\text{post}$ is traversed by inflation, which implies that $a_e/c > c a_i$, or, equivalently, that $c^2 < e^{N_t}$.  Hence, the parameters set we work with is $(r,c)$.

Again, it proves convenient to rewrite~(\ref{cAmod2}) in terms of $N$ as
\be\label{cAmod2N}
\cA_h &=& \frac{1}{\sqrt{2k}} \left\{ \left[ \cth_\text{pre} + e^{r(N - N_\text{pre})} W + e^{r(N_\text{post} - N_\text{pre})} \cth_\text{post} \right] \Theta \right. \nonumber\\
&&+ \left. \left[ e^{r N} W + e^{r N_\text{post}} \cth_\text{post} \right] \bar\Theta \right\} e^{-ik\eta} \, ,
\ee
from which we can compute the power spectrum as
\be\label{deltaBmod2}
\delta_B^{2,e} &=& \frac{H_I^4}{8\pi^2} \left\{ c^{2r} \left( \frac{k}{\cH_e} \right)^{4 - 2r} W^e \Theta^e + c^{4r} \left( \frac{k}{\cH_e} \right)^4 \cth_\text{post}^e \Theta^e \right.\nonumber\\
&& + \left. c^{2r} \left( \frac{k}{\cH_e} \right)^{4 + 2r} e^{2rN_t} \cth_\text{post}^e \bar\Theta^e \right\} \, ,
\ee
where the first term refers to high energy modes for which the amplification window was not traversed completely due to the end of inflation, the second term encompasses all modes which have been fully amplified, and the last, third, term, describes low-energy modes which were being amplified right at the beginning of inflation.  The final energy density is given by
\be\label{energymod2}
\rho_\gamma^e &=& \frac{H_I^4}{8\pi^2} \left\{ \frac{1}{2-r} \left[ c^{2r} - c^{4(r-1)} \right] + \frac12 \left[ c^{4(r-1)} - c^{4(r+1)} e^{-4N_t} \right]  \right. \nonumber\\
&& + \left. \frac{r^2}{2-2r} \left[ c^{2r} - c^{2(2r-1)} \right] + \frac{1}{2+r} \left[ c^{4(r+1)} - c^{2r} \right] e^{-4N_t} \right\} \, ,
\ee
where the magnetic field energy density is made of half the first, second, and last term, the rest pertaining to the electric field.  We can understand the first piece as belonging to the highest energy scales, which were not fully amplified, the second piece represents all intermediate scales with full amplification, the third is similar to the first but arises from the conformal time derivative in the electric field, and the last one accounts for the lowest energy modes which were also not fully amplified.  Notice that the result is barely dependent on the total number of e-foldings as it should be, as the width of the window is independent on the dynamics of inflation.

\paragraph{Model 3.} The parametrisations described so far are not knowledgeable of the mode being processed, except for its being in, out, or around the time of horizon crossing.  This is normally the case as the coupling functions are independent of $k$ itself.  However, one can envisage a situation where this restriction is dropped, and the efficiency at which a mode is amplified does depend directly on the mode itself.  Since the typical amplification mechanism is operative outside the Hubble scale, we choose to allow for a $k$-dependence in the exponent $q$, which becomes now $\hat q \equiv q (k/\hat k)^d$, and where the reference scale $\hat k$ is fixed at 1/Mpc, for this is the characteristic scale we will explore numerically.

The parameters set thus becomes $(q,d)$, with $d$ brought in to define the ``tilt'' of the amplification mechanism; positive values for $d$ mean that the higher end of the spectrum will have a higher boost factor compared to the red end, while a negative $d$ would favour large scales over blue ones.  Explicitly:
\be\label{cAmod3}
\cA_h = \frac{1}{\sqrt{2k}} \left\{ \cth_\text{in} + \left( \frac{a}{a_k} \right)^{\hat q} \cth_\text{out} \right\} e^{-ik\eta} \, ,
\ee
or, in terms of e-foldings,
\be\label{cAmod3N}
\cA_h = \frac{1}{\sqrt{2k}} \left\{ \cth_\text{in} + |k\eta|^{- \hat q} \cth_\text{out} \right\} e^{-ik\eta} \, .
\ee

Once again, the final power spectrum at the end of inflation is
\be\label{deltaBmod3}
\delta_B^{2,e} = \frac{H_I^4}{8\pi^2} \left( \frac{k}{\cH_e} \right)^{4 - 2 \hat q} \, ,
\ee
and the total electric and magnetic energy density is the integral
\be\label{energymod3}
\rho_\gamma^e = \frac{H_I^4}{8\pi^2} \int \dd\kappa \left\{ 2\kappa^3 + \kappa \hat q^2 \right\} \kappa^{2 \hat q} \, ,
\ee
where we have defined $\kappa \equiv k/\cH_e$ for convenience.  It is in general not possible to solve~(\ref{energymod3}) in a closed form, due to the $k$-dependent exponent $\hat q$, so we will limit ourselves to directly solve the integral by numerical means.

\paragraph{Model 4.} This parametrisation deals with the issue of explicit cutoffs in the magnetic (and electric) field spectrum.  So far one further, unspoken, but basic assumption has been that all modes from $k_\text{min}$ up to $k_\text{max}$ partake in the amplification process.  As is clear already by peeking at the simplest results Eqs.~(\ref{energymod1}) and~(\ref{energymod2}), the highest scales are very likely to be dominant in the energy budget of the electromagnetic field, and the most dangerous in terms of their possible backreactions.  The fact that the total energy density has to be shared by such a wide range of modes makes it much harder for single modes, especially large scale modes, to be entitled to any relevant portion of it.  One may thence hope that by cutting off part of the spectrum, because of some physical cutoff in the theory which produces the $\cL_{\cph\gamma}$ interaction term, the situation could be ameliorated.

The range of modes which appear to be the trickiest to generate has been already identified in the low energy one, at and around the Mpc scale.  Thus, of the two possible cutoffs, the infrared and the ultraviolet one, the most interesting would be the latter, for there are only a handful of modes below 1/Mpc which could be removed.  Physically speaking, it also seems reasonable to work with an ultraviolet cutoff for its connection to the effective field theory producing the higher order coupling terms in the Lagrangian. While we will retain the $p=0$ assumption in this class of models, we will discuss both cutoffs in what follows, in connection to their different physical interpretation.

The high energy cutoff in the theory is implemented by simply restricting the range of integration in~(\ref{energy}), and attaching a Heaviside step to the power spectrum~(\ref{deltaB}) as
\be\label{deltaBmod4}
\delta_B^{2,e} = \frac{H_I^4}{8\pi^2} \left( \frac{k}{\cH_e} \right)^{4 - 2q} \cth_\text{uv}\, ,
\ee
and
\be\label{energymod4}
\rho_\gamma^e = \frac{H_I^4}{8\pi^2} \left\{ \frac{1}{2-q} c^{(4-2q)} \left[ 1 - e^{(2q - 4)N_t} \right] + \frac{q^2}{2-2q} c^{2-2q} \left[ 1 - e^{(2q - 2)N_t} \right] \right\} \, ,
\ee
where we have used $0<c<1$ to set the upper limit for $k$ through $\cth_\text{uv} \equiv \cth(c \cH_e - k)$.  The parameters set is $(q,c)$, however, these expressions are easily generalisable to include within-horizon amplification (through $p$), or accommodate an infrared cutoff (through $\cth_\text{ir} \equiv \cth(c k - \cH_i)$): in the latter case the final energy density~(\ref{energymod4}) is modified as
\be\label{energymod5}
\rho_\gamma^e = \frac{H_I^4}{8\pi^2} \left\{ \frac{1}{2-q} \left[ 1 - c^{-(2q - 4)}e^{(2q - 4)N_t} \right] + \frac{q^2}{2-2q} \left[ 1 - c^{-(2q - 2)} e^{(2q - 2)N_t} \right] \right\} \, .
\ee

\paragraph{Model e.} In this last example we want to see what happens in the extreme case of an exponential growth for $\cA_h$.  The relevance of this possibility lies in the fact that it is perhaps the most immediate and direct way to put electric and magnetic field on the same level, if not even to suppress the former compared to the latter.  The details of the struggle for power of the two will be analysed below, so for now let us simply introduce the parameters $(p,q)$ or $(P,Q) \equiv (p\omega,q\omega)$ as
\be\label{cAmodE}
\cA_h = \frac{1}{\sqrt{2k}} \left\{ e^{p\omega(\eta-\eta_i)} \cth_\text{in} + e^{p\omega(\eta_k-\eta_i)+q\omega(\eta-\eta_k)} \cth_\text{out} \right\} e^{-ik\eta} \, ,
\ee
where we can think of the frequency $\omega$ as exactly $k$ or some independent parameter (thence the re-parametrisation $p\omega \rar P$ and $q\omega \rar Q$).

For the last time, let us also explicitly write down the final power spectrum~(\ref{deltaB})
\be\label{deltaBmodE}
\delta_B^{2,e} = \frac{H_I^4}{8\pi^2} \left( \frac{k}{\cH_e} \right)^4 e^{2P(\eta_k-\eta_i)+2Q(\eta_e-\eta_k)} \, ,
\ee
where $\eta_k \equiv -1/k$ is negligible compared to $\eta_i$, and $\eta_e$ so is for $\eta_k$:
\be\label{deltaBmodEs}
\delta_B^{2,0} \simeq \frac{k^4}{8\pi^2} e^{2p \left( \frac{\omega}{\cH_i} \right) + 2q \left( \frac{\omega}{k} \right) } \, ;
\ee
the total electromagnetic energy density~(\ref{energy}) reads
\be\label{energymodE}
\rho_\gamma^e = \frac{H_I^4}{8\pi^2} \int \dd\kappa \left\{ 2\kappa^3 + \kappa \left( \frac{Q}{\cH_e} \right)^2 \right\} e^{2P(\eta_k-\eta_i)+2Q(\eta_e-\eta_k)} \, .
\ee
The electric field contribution here, besides the usual first half of the first term in the curly brackets of~(\ref{energymodE}) which matches that of the magnetic field, is proportional to the square of $q\omega/\cH_e$, and will depend on the specific values for $\omega$ and $q$.  This expression can be simplified by introducing $\hat Q \equiv Q / \cH_e$, $\hat P \equiv P / \cH_e$, and ignoring the $|\eta_k| \ll |\eta_i|$ and $|\eta_e| \ll |\eta_k|$ in the exponents, to give
\be\label{energymodEs}
\rho_\gamma^e &\simeq& \frac{H_I^4}{8\pi^2} \int \dd\kappa \left\{ 2\kappa^3 + \kappa \hat Q^2 \right\} e^{2 \hat P e^{N_t} + 2 \hat Q / \kappa} \nonumber\\
&=& \frac{H_I^4}{8\pi^2} e^{2 \hat P e^{N_t}} \left. \left\{ \frac16 e^{2 \hat Q / \kappa} \left[ 3\kappa^4 + 2\kappa^3 \hat Q + 5\kappa^2 \hat Q^2 + 10\kappa \hat Q^3 \right] - \frac{10}{3} \hat Q^4 \text{Ei} (2 \hat Q / \kappa) \right\} \right|_{e^{-N_t}}^1 \nonumber\\
&\simeq&
\frac{H_I^4}{8\pi^2} \frac12 e^{2 p \left( \frac{\omega}{\cH_i} \right) } \left\{ 1 + q \left( \frac{\omega}{\cH_i} \right) e^{2 q \left( \frac{\omega}{\cH_i} \right) - 4N_t} \right\} \, .
\ee
The last expression is technically valid for values of $Q = q\omega \gtrsim \cH_i$, but due to the extra factor of $\exp(-4N_t)$ in the second term, whenever $Q \lesssim \cH_i$ the last term becomes completely insignificant anyhow.

\subsection{Background dynamics}\label{INFevol}

The background evolution is determined by the details of the inflationary expansion epoch, which we characterise in this section.  Although we will stick to a very simple and circumstantiated example (mostly chaotic inflation), the results we obtain do apply generally to other types of inflation; in the end the only parameter which makes the difference is the scale of inflation, which sets the ``distance'' between the end of inflation to today, thereby determining the dilution factor for the magnetic energy density and strength at given wavelength.

The background dynamics is governed by a scalar field $\cph$, the inflaton, with potential $V(\cph)$.  In order to be definite, and in order to be able to fix all parameters for the subsequent numerical calculations, we choose to work with either a monomial power-law potential or the type
\be\label{infV}
V(\cph) = M^4 \left( \frac{\cph}{M} \right)^\alpha \, , \text{ for } \, \alpha\neq4 \quad \text{ or } \quad V(\cph) = \lambda \cph^4 \, \text{ for } \, \alpha=4 \, ,
\ee
where the mass scale $M$ (or the coupling constant $\lambda$) determines the scale of inflation; we will also discuss briefly one example of small-field inflation, exemplified by the potential
\be\label{infVv}
V(\cph) = M^4 (1 + \cos g\cph) \, ,
\ee
where the dimensionful coupling $g$ is the inverse ``decay constant'' for a would-be (pseudo) Nambu-Goldstone boson ($\cph$ in this case).

Given the potential one can readily obtain the dynamics, assuming there are no backreactions in effect:
\be\label{infEOM}
\ddot\cph + 3H\dot\cph + V_\cph = 0 \, ,
\ee
where $V_\cph \equiv \dd V/\dd\cph$.  The inflaton will start somewhere up high in the potential, and slowly roll down until its kinetic energy density, which is assumed to be unimportant at this stage, catches up with the decaying vacuum potential energy, and finally causes the inflaton to decay while rapidly oscillating around its minimum.  A handy way to follow this evolution is through the number of e-foldings for which inflation lasts, which we can approximate as
\be\label{efolds}
N \simeq \frac{1}{M_4^2}\int \frac{V}{V_\cph} \dd\cph \, .
\ee

The end of inflation is typically estimated to happen when the slow-roll parameters
\be\label{slowroll}
\varepsilon_{SR} \equiv - \frac{\dot{H}}{H^2} \simeq \frac{M_4^2}{2} \left(\frac{V_\cph}{V}\right)^2 \quad,\quad \eta_{SR} \equiv 2\varepsilon_{SR} - \frac{\dot{\varepsilon}_{SR}}{2H\varepsilon_{SR}} \simeq M_4^2 \left(\frac{V_{\cph\cph}}{V}\right) \, ,
\ee
become of order one, hence $\cph_f \approx \alpha M_4 / \sqrt2$ (similar results hold for the $\alpha=4$ case).  This means that the evolution of $\cph$ itself is
\be\label{infevol}
(\frac{\cph}{M_4})^2 \approx  \frac{\alpha^2}{2} + 2\alpha N \, .
\ee

The scale of inflation is determined by $M$, which ultimately will yield $H_I$, wherefrom we can make the connection to our Universe by assuming the type of evolution (matter radiation, or else) in the wake of the aftermath of the de Sitter expansion.  This parameter is fixed, for a given $N_t$ by normalising the generated curvature perturbations to the directly related temperature fluctuations as measured by CMB experiments such as WMAP.  The power spectrum of such temperature fluctuations is encoded in the quantity
\be\label{deltaH}
\delta_H^2 = \frac{1}{150\pi^2 M_4^4} \frac{V(\cph_i)}{\varepsilon_{SR}(\cph_i)} \approx (2\times10^{-5})^2 \, ,
\ee
where the field $\cph$ is evaluated at the end of inflation.  Using this result, one can fix the scale of inflation $H_I^2 \equiv V/3M_4^2$ as
\be\label{HI}
H_I^2 = \frac{50\pi^2}{(1 + 4N_t / \alpha)^{\alpha/2 +1}} \delta_H^2 M_4^2 \, .
\ee
One more piece of information which we will be using, effectively as a way to parametrise variations in the total life span of inflation, is the spectral index of the aforementioned perturbations, which for monomial inflation reads
\be\label{ns}
n_s = 1 - \frac{2(\alpha+2)}{\alpha + 4N_t} \, .
\ee

This slightly non-standard take on inflation is adopted for the ease of numerical implementation which thereby comes about; the results we will obtain and document do not crucially depend on these details however, and can be straightforwardly adapted and generalised.  In our analysis we will always work with the total $N_t$ also because we do not want to descend too deeply in the swampy lands of actual quantisation and renormalisation of the electromagnetic potential (and in turn of the inflaton field~\cite{Finelli:2011cw,Marozzi:2011da}), which will take us far beyond the scope of this paper.  We will hence be content with always ``booking'' a range of modes going from $k_\text{min}$ to $k_\text{max}$ which depends on $N_t$ in its width, and on $H_I$ (which can or can not be constrained given $N_t$) for its absolute value; clearly, one needs to ensure, a posteriori, that the modes we are looking at are within the range touched by inflation.

\section{How strong can the magnetic fields possibly be?}\label{results}

The key quantity which roughly determines the onset of backreactions is the ratio
\be\label{ratio}
x = \frac{\rho_\gamma}{\rho_\cph} \, ;
\ee
when this ratio becomes of order one (being negligible at the beginning of inflation) the background becomes unstable, since the Friedmann equation $H^2 \simeq \rho_\cph + \rho_\gamma$ does not admit quasi de Sitter solutions anymore.  Obviously we do not really know what happens when $x\rar1$ until we solve the coupled Euler-Lagrange equations for $\cph$ and $\cA_h$, but it seems safe to assume that, if inflation surpasses the impediment, the final magnetic field will be negligible, see~\cite{Demozzi:2009fu,Kanno:2009ei,Emami:2009vd,Durrer:2010mq}~\footnote{Notice that in the specific model of~\cite{Campanelli:2008kh} the problem does not arise because only large scales are touched, \emph{and} the amplification takes place while these are superhorizon: as we find here, this might be the only viable combination.}.

The time at which the $\cA_h$ backreacts could be found instead by directly looking at the equations of motion~(\ref{infEOM}) where terms explicitly containing $I_\cph(\eta)$, $f_\cph(\eta)$, or $m(\eta)$ appear multiplying some combination of $\cA_h$.  The balance is then between $V_\cph$ and the latter, and can in principle be very different from $x$.  However, to investigate this we need to know explicitly the form of the interaction term, so we do not pursue this possibility any further, bearing in mind that our results are a conservative estimate, in the sense that the upper limits we will derive could be further lowered.

\subsection{Finding the maximal allowed strength}\label{fixNgetB}

Let us pick a particular background and look at the limits on the model parameters.  To begin with, we will choose to fix the spectral index of~(\ref{ns}) to $n_s = 0.97$ and normalise the density fluctuations $\delta_H$ as in~(\ref{deltaH}).  This for a given potential, say chaotic inflation with a quadratic potential $\alpha=2$, provides us with the total duration of inflation, in this case around $N_t \approx 66$.  The only unknowns are then the model parameters sets: $(p,q)$, $(r,c)$, $(q,d)$, and $(q,c)$, for the four models we have introduced above.

\begin{figure}[ht]
\centering
\includegraphics[width=0.48\textwidth]{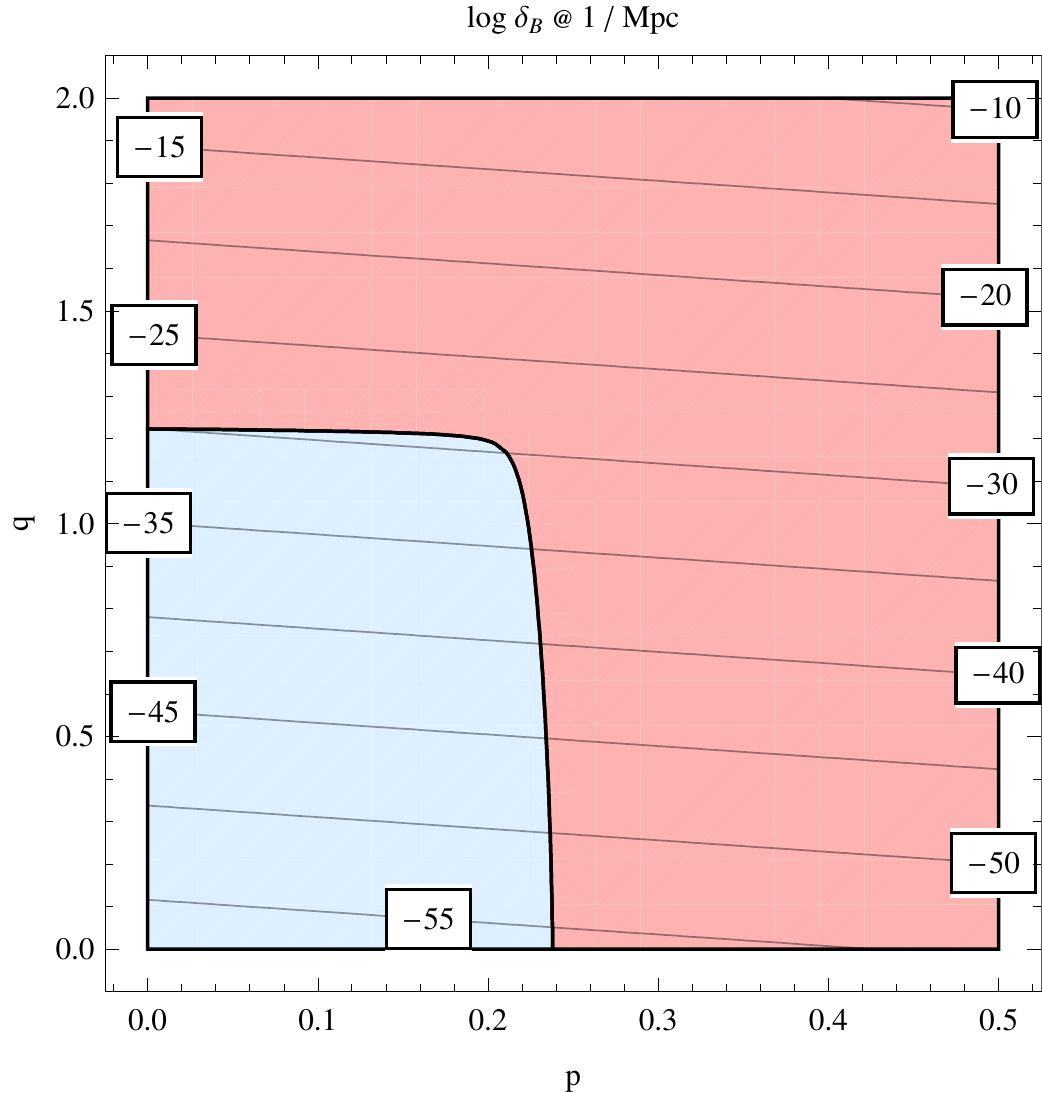}
\caption{Allowed (in light blue) and disallowed (in light red) regions in the $(p,q)$ parameters space.  Superimposed are contours of constant $\log\delta_B^0$ today, at 1/Mpc (in Gauss units).}
\label{density1}
\end{figure}

\begin{figure}[ht]
\centering
\includegraphics[width=0.48\textwidth]{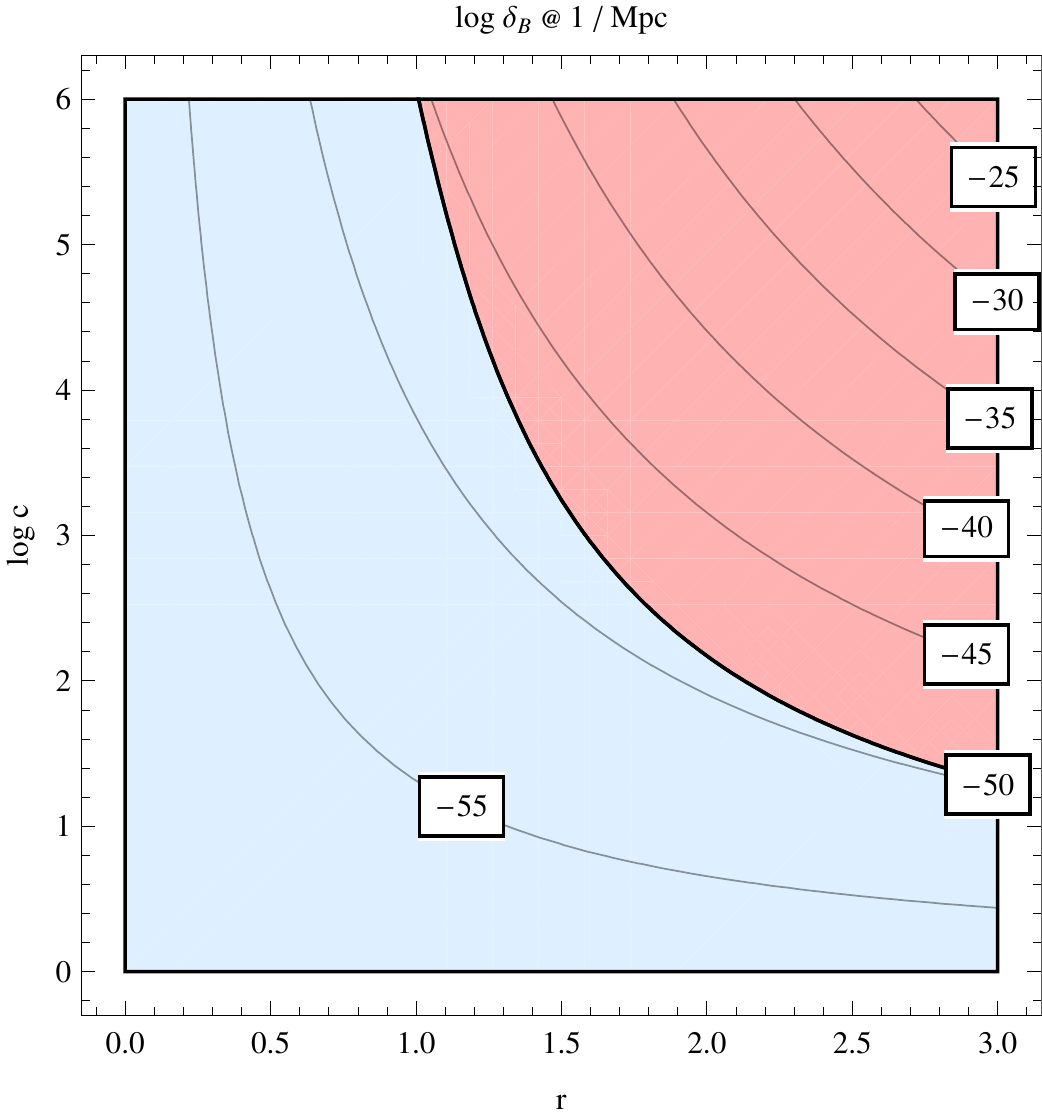}
\caption{Allowed (in light blue) and disallowed (in light red) regions in the $(r,\log c)$ parameters space.  Superimposed are contours of constant $\log\delta_B^0$ today, at 1/Mpc (in Gauss units).}
\label{density2}
\end{figure}

\begin{figure}[ht]
\centering
\includegraphics[width=0.48\textwidth]{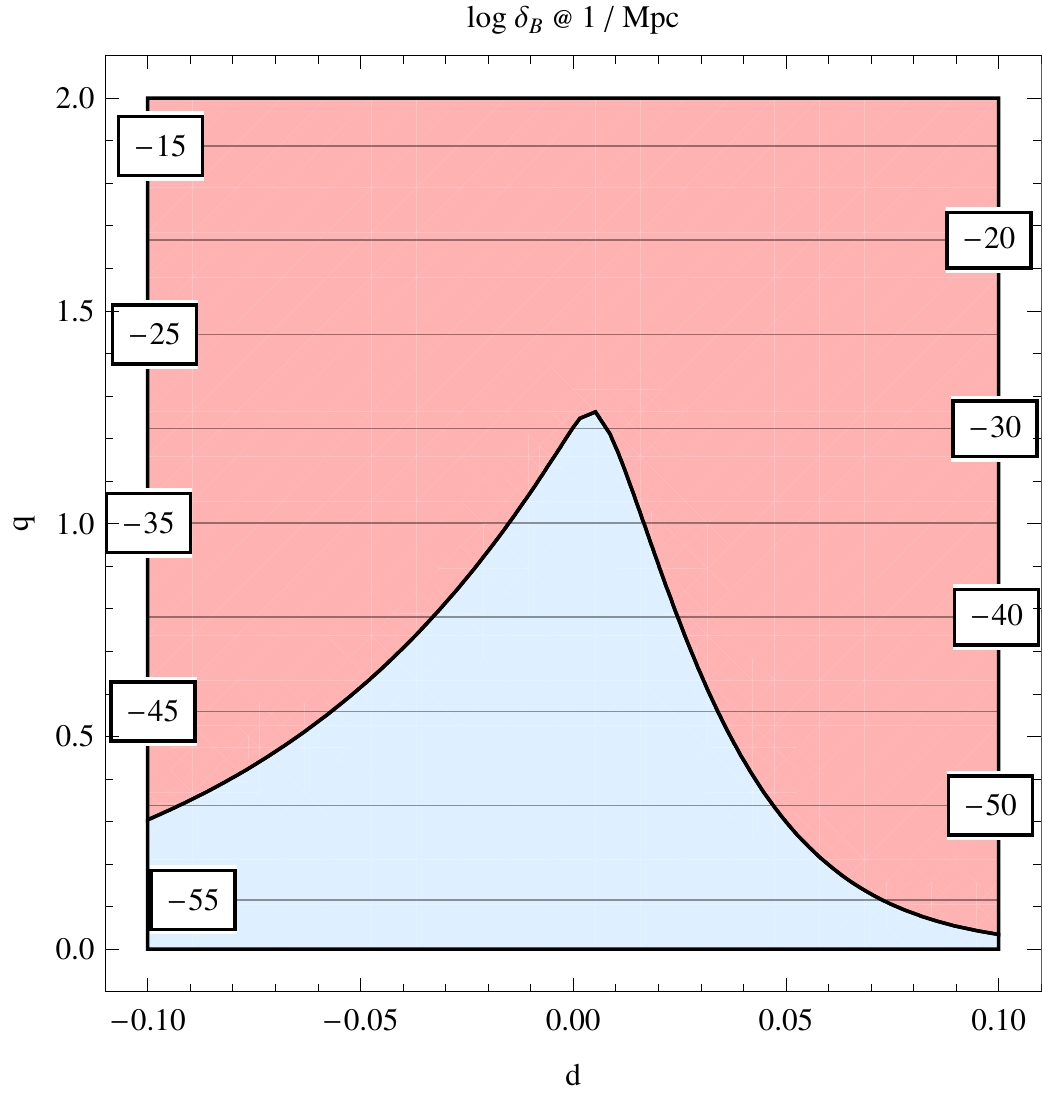}
\caption{Allowed (in light blue) and disallowed (in light red) regions in the $(d,q)$ parameters space.  Superimposed are contours of constant $\log\delta_B^0$ today, at 1/Mpc (in Gauss units).}
\label{density3}
\end{figure}

\begin{figure}[hb]
\centering
\includegraphics[width=0.48\textwidth]{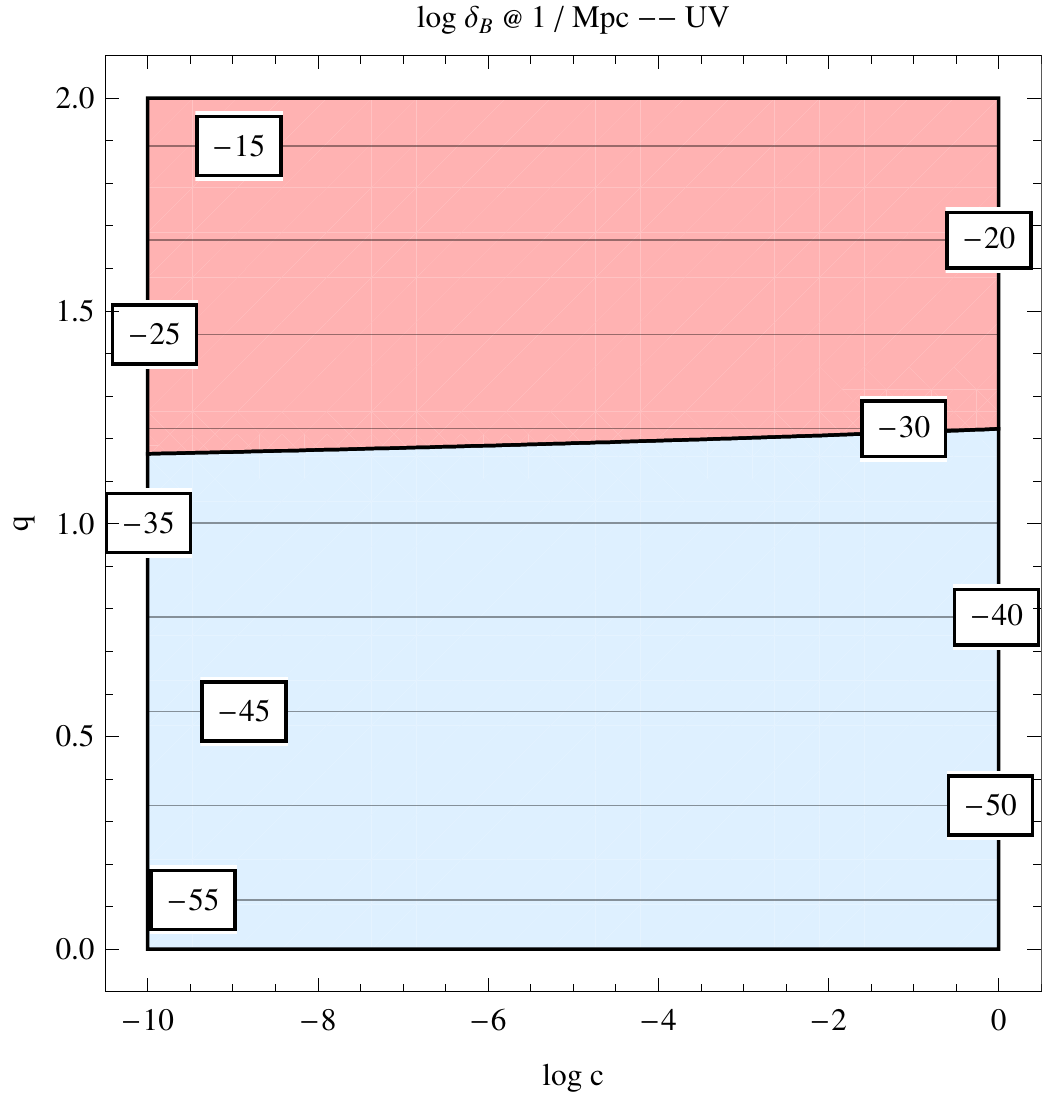}\hspace{3mm}
\includegraphics[width=0.48\textwidth]{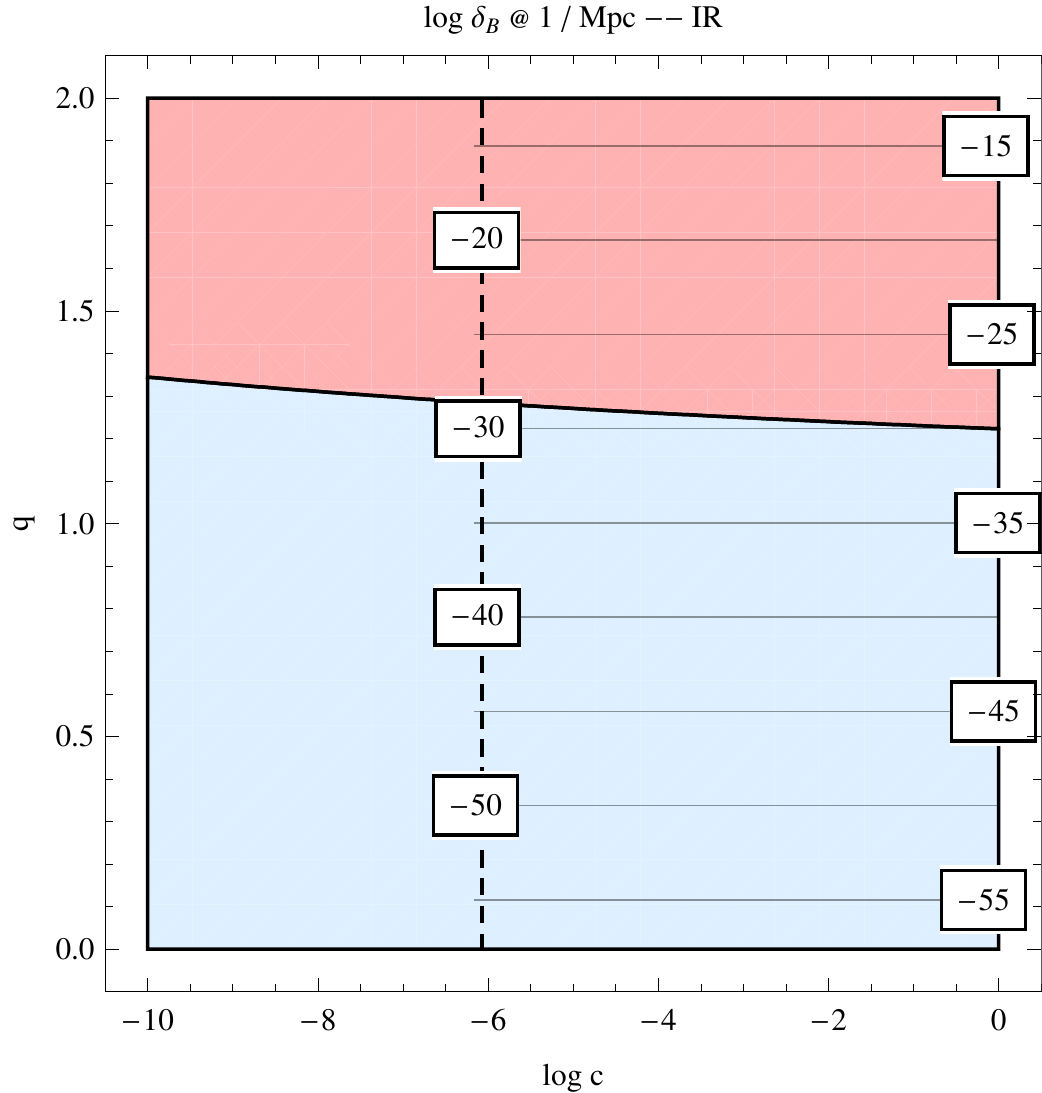}
\caption{Allowed (in light blue) and disallowed (in light red) regions in the $(\log c,q)$ parameters space for the ultraviolet (left panel) and infrared (right panel) cutoffs.  Superimposed are contours of constant $\log\delta_B^0$ today, at 1/Mpc (in Gauss units).  The dashed line in the right panel marks the value of $c$ for which 1 Mpc is cut off; no power on this scale is generated for values of $c$ below this line.}
\label{density4}
\end{figure}

We look at regions in the parameters space where the ratio~(\ref{ratio}) stays below one, which are plotted in Figs.~\ref{density1},~\ref{density2},~\ref{density3}, and~\ref{density4}; on top of the allowed and disallowed regions (according to this criterion) we draw the contours representing (the logarithm of) the strength of the magnetic field $\delta_B^0$ today.  Notice that, again for the sake of simplicity, we assume that the Universe reheats perfectly and instantaneously, delivering a beautiful radiation-filled primordial soup; this is expressible as
\be\label{connect}
\rho_\cph^e \simeq T_e^4 \, , \quad a_e = \frac{T_0}{T_e} a_0 \, ,
\ee
where $a_0=1$, and $T_0 = 2.75$ K is today's cosmic microwave background's temperature -- we are disregarding the contribution of neutrinos, as well as the latter dark energy.

\paragraph{Model 1.} The first parametrisation, Eq.~(\ref{cAmod1N}) plainly describes a simple growth inside and outside the horizon at different rates $p$ and $q$, respectively.  The allowed regions and the generated magnetic field strength are shown in Fig.~\ref{density1}.  Clearly smaller values of these parameters will be required for a long enough inflation.  Notice that the parameter $q$ can never be bigger than approximately 1.2, meaning that the magnetic field spectrum will always be blue.

\paragraph{Model 2.} The second parameters set, $(r,c)$ from~(\ref{cAmod2N}), accounts for a fixed-time growth around horizon crossing (as it typical in models where photons and inflation are axially coupled).  As expected, see Fig.~\ref{density2}, large $r$ are not allowed, especially as we widen the window with $c$, since in that case the modes would be amplified for longer, and backreact more easily.  We have also noticed already that this result depends only very little on the total number of e-foldings.  The spectrum is typically blue, although it is possible to have a red ``tail'' for high energy modes: the large scale modes are in any case very blue.

\begin{figure}[hb]
\centering
\includegraphics[width=0.48\textwidth]{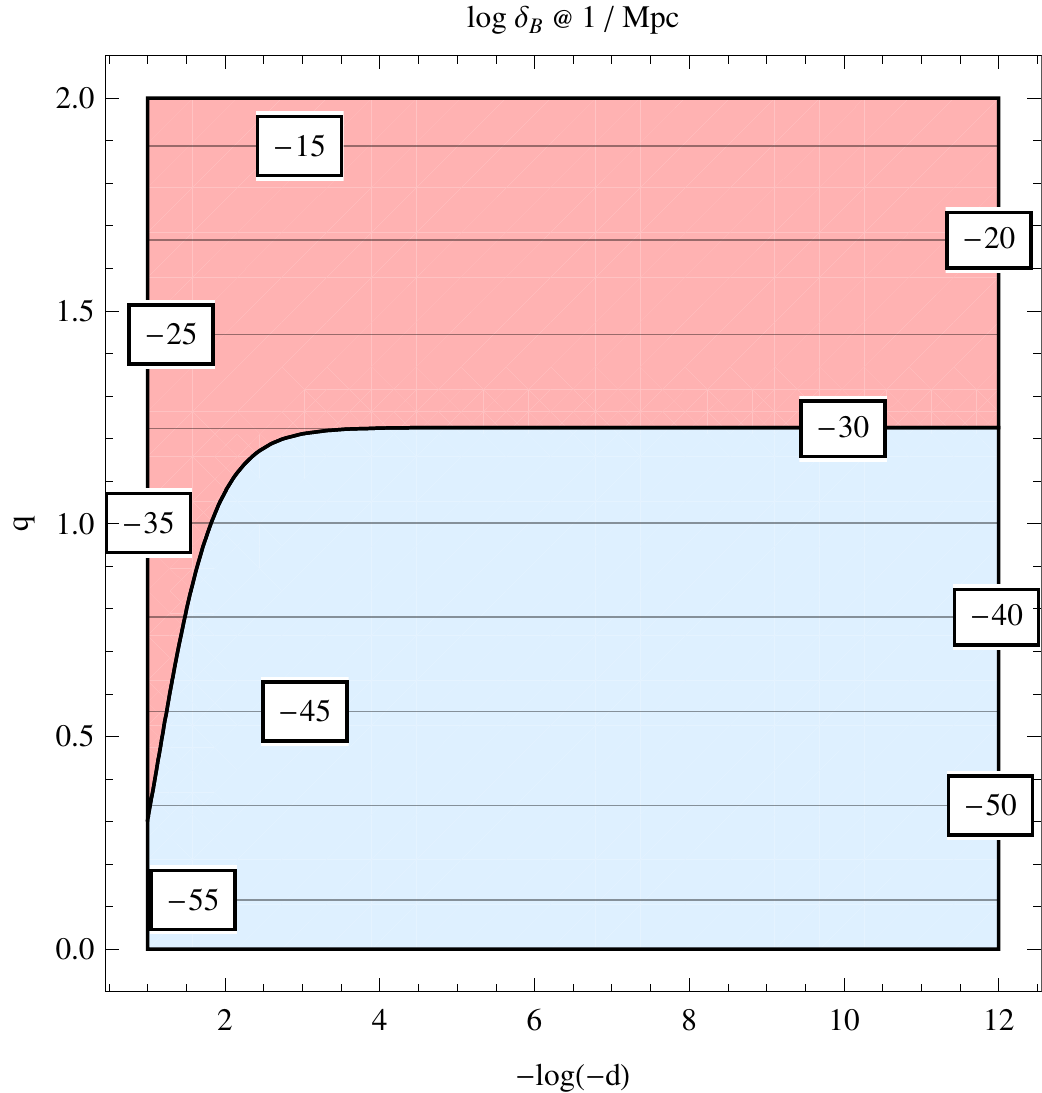}\hspace{3mm}
\includegraphics[width=0.48\textwidth]{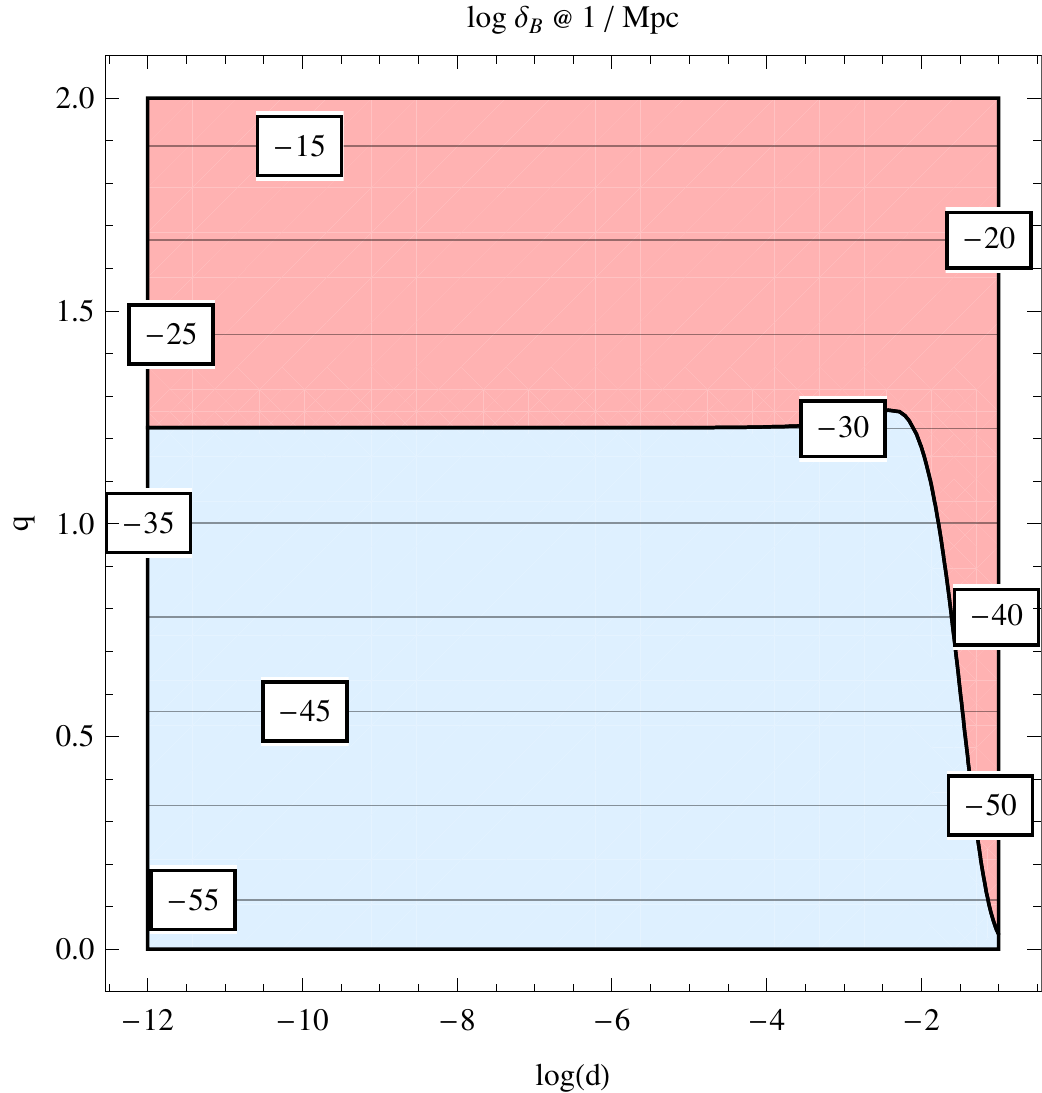}
\caption{Zoom on Fig.~\ref{density3} for small negative (left panel) and positive (right panel) values of $d$, expressed in terms of $-\log(-d)$ and $\log(d)$, respectively.}
\label{density3zoom}
\end{figure}

\paragraph{Model 3.} The running of the amplification factor with $k$ is expressed in~(\ref{cAmod3N}) through the set $(q,d)$.  Notice (Fig.~\ref{density3}) that the largest magnetic field strength on large scales is obtained for $d \simeq 0$, meaning that uniform amplification is preferred.  We can understand this by recalling that the range of modes in question is numerically huge, as $\kappa \in [e^{-N_t},\,1]$, and even relatively small values for $d$ will enhance dramatically one or the other end of the spectrum, thereby tightening the constraints on the growth parameter $q$.  Looking at it in a logarithmic scale in Fig.~\ref{density3zoom} we see that very moderate values for $d$ in fact do ameliorate the situation, but clearly not enough to allow for a significant magnetic field today.

\paragraph{Model 4.} The two following examples refer to Eqs.~(\ref{energymod4}), and~(\ref{energymod5}), Fig~\ref{density4}, left and right panels, for the ultraviolet and infrared cutoffs, respectively.  In the first case the change is minimal, because once again, due to the different powers for the electric and magnetic contributions, as soon as $q>1$ the electric spectrum turns red, and the high energy cutoff becomes uninfluential.  In the opposite case instead, where we cut off large scale modes, there can be seen a fairly more significant difference; nonetheless, this possibility is still not viable, because there are only a handful of modes which can be cut off before reaching 1 Mpc, and at this scale the electric field is still too strong, and redder, than the magnetic field.

\paragraph{Model e.} Speaking of the relevance of the electric field, as will be sketched in the next section, one simple way to prevent it from growing much faster than the magnetic field is to engineer some exponential-type of amplification for $\cA_h$; this is what was proposed in~(\ref{cAmodE}).  The spectrum~(\ref{deltaBmodEs}) is blue unless the exponential tilts it; in particular, the lowest energy mode $k_\text{min} = \cH_i$ has the same power of the highest $k_\text{max} = \cH_e$ (or more) if $2 q \omega / k_\text{min} \geq 4 N_t$, even though it decays exponentially for higher modes, until the $k^4$ overcomes it.  A closer look at Eq.~(\ref{energymodEs}) reveals that as soon as the spectrum ``turns red'', the electric contribution becomes dominant once again, because of the $q\omega/\cH_i$ prefactor, which is large in this case: the electric field rears its ugly head again.  Despite the ratio $x$ being still well below one at the turning point for $q\omega$, it climbs up exponentially rapidly, and does not allow for more than about $10^{-58}$ Gauss today on scales of a Mpc.

\subsection{Variations on a theme}\label{fixBgetN}

There are several variations on this work which can be investigated; among the several possibilities available we focus on a few examples.

\begin{figure}[ht]
\centering
\includegraphics[width=0.48\textwidth]{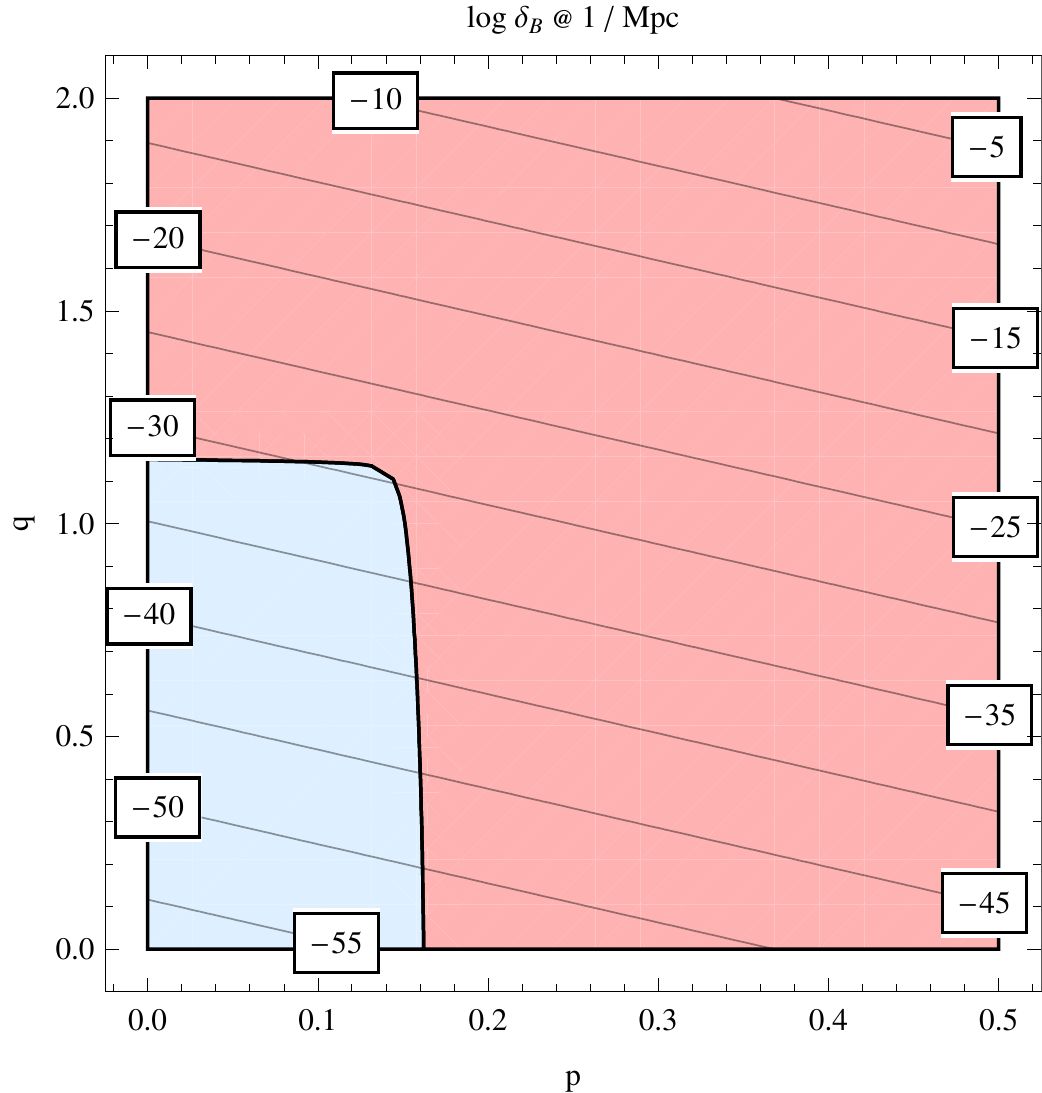}\hspace{3mm}
\includegraphics[width=0.48\textwidth]{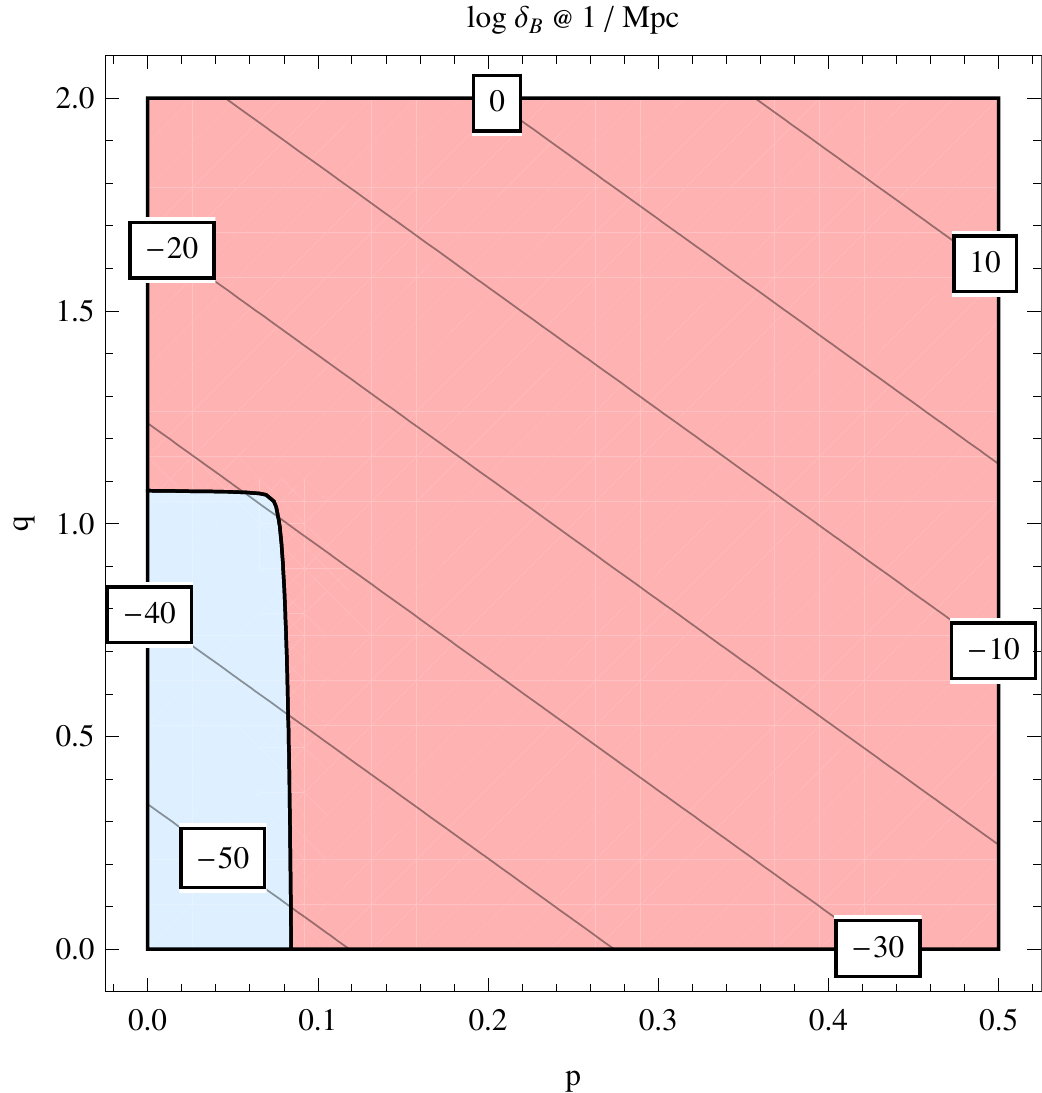}
\caption{Same as Fig.~\ref{density1}, but for $n_s = 0.98$ (left panel) and $n_s = 0.99$ (right panel); this changes the overall duration of inflation to $N_t = 99.5$ and $N_t = 199.5$, respectively.}
\label{densityns1}
\end{figure}

\begin{figure}[hb]
\centering
\includegraphics[width=0.48\textwidth]{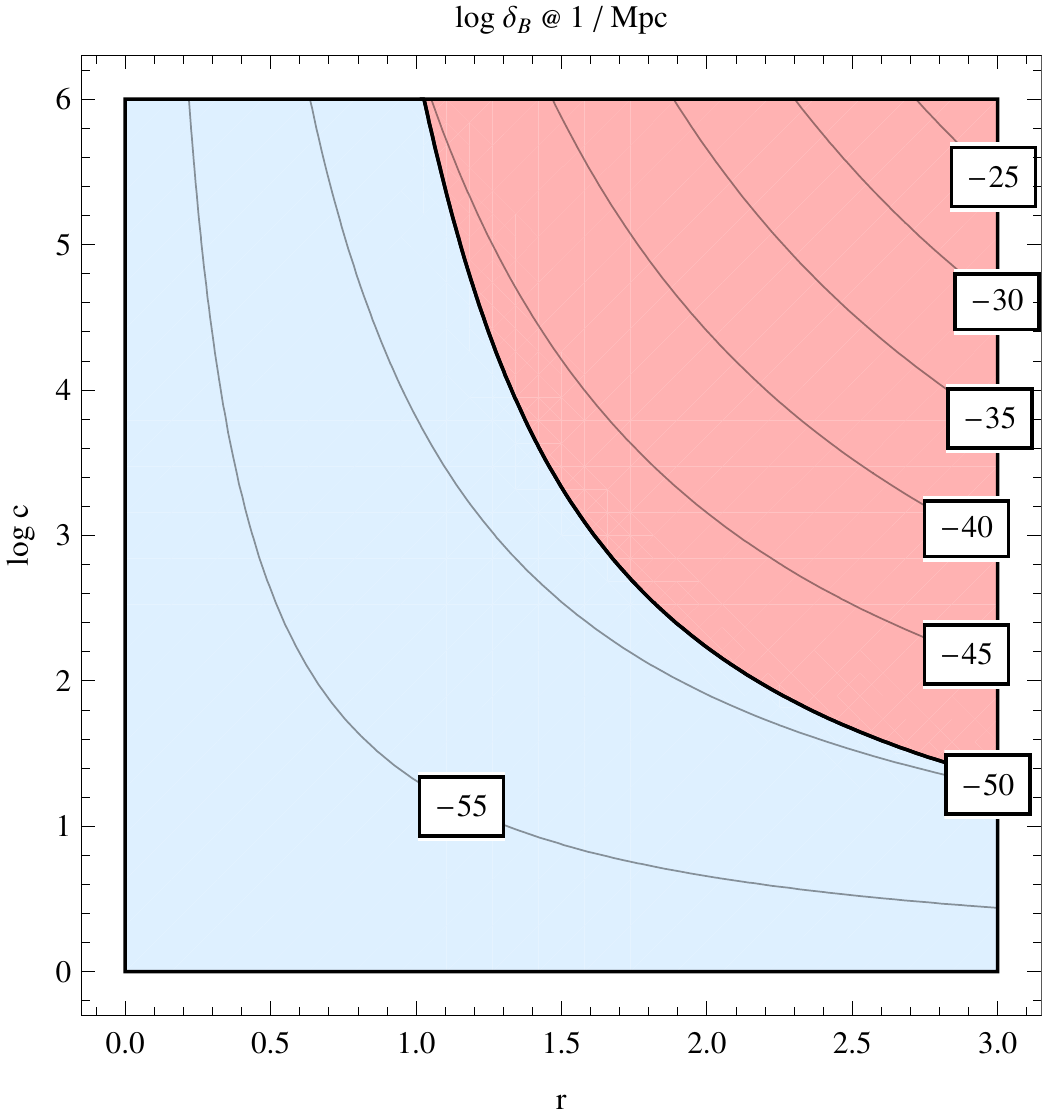}\hspace{3mm}
\includegraphics[width=0.48\textwidth]{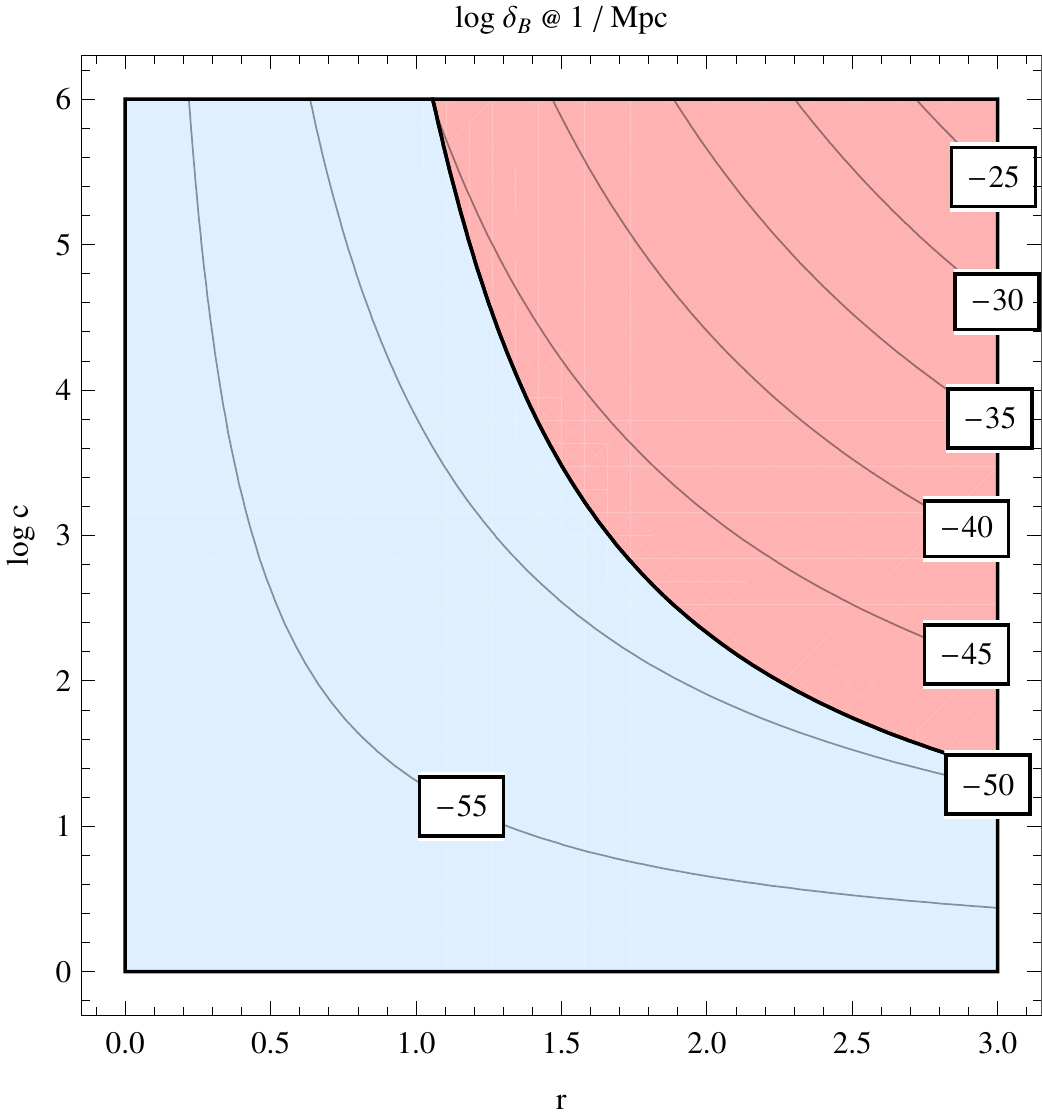}
\caption{Same as Fig.~\ref{density2}, but for $n_s = 0.98$ (left panel) and $n_s = 0.99$ (right panel); this changes the overall duration of inflation to $N_t = 99.5$ and $N_t = 199.5$, respectively.}
\label{densityns2}
\end{figure}

\paragraph{Varying $n_s$.} First, we can vary the scalar index $n_s$, which effectively is a way to play with the total number of e-foldings.  Of course, as long as we stick to the simple quadratic potential, and given that we need at least around 60 e-foldings to be able to let today's Hubble scale $\cH_0$ be inside the inflating machine once, there is only that much freedom in choosing $n_s$.  In Figs.~\ref{densityns1}, and~\ref{densityns2} we compare the allowed regions where $x<1$ where we rise $n_s$ to 0.98 and 0.99, respectively.  We do so for the first two models with sets $(p,q)$ and $(r,c)$, respectively; in the first case the ``good'' parameters space further shrinks as expected -- more inflation means more growth; in the last case the results are barely dependent on $N_t$, if not for a slight widening of the blue region.

\paragraph{Varying $V(\cph)$.} Next, we can tamper with the inflaton potential with the hope of finding different relations between the background parameters $n_s$, $H_I$, and $N_t$, which could help in alleviating the severe dilution the generated magnetic field incurs in.  Since we have not found any significant departure from the already reported outcomes by simply switching to quartic or other simple monomial potential, we turn our attention to the ``natural'' inflation case, Eq.~(\ref{infVv}).

In natural inflation the scalar index turns out to be directly linked to the inflaton's inverse decay constant $g$ by
\be\label{nsv}
n_s = 1 - (gM_4)^2 \, ,
\ee
which means that if observations pin down the value for $n_s$, we automatically fix $g$, but we still have complete freedom as for $N_t$, unlike the case of a monomial potential, see~(\ref{ns}).  The most important consequence of this is reflected in the final $H_I$ which now explicitly depends on $n_s$ and $N_t$ separately:
\be\label{HIv}
H_I^2 \simeq \frac{75\pi^2}{4} \delta_H^2 e^{(n_s -1)N_t} M_4^2 \, ,
\ee
up to a factor of order unity, weakly dependent on $n_s$ itself.  However, with typical values for the scalar index and the total e-foldings, also~(\ref{HIv}) falls in the usual range of around and above $10^{13}$ GeV, and no significant changes occur.

\paragraph{Flipping the argument.} One can embrace the somewhat backwards approach of deciding first what the desired magnetic field strength is, and given that compute the allowed duration of inflation for given set of parameters.  For instance, we can agree to succumb to the observational imperative demanding nGauss strength on scales of a Mpc, and work out how long a successful inflation could possibly last.  This point of view is in fact a simple ``re-normalisation'' of the results of the previous sections, for instead of asking how large will $\delta_B$ be with the constraint $x=1$, we ask how long will $x<1$ be satisfied for given the final $\delta_B$, and we will not pursue it further.

\subsection{Some physics}\label{some}

The results of the previous section show how undoubtedly challenging it is to generate magnetic fields of tangible intensity today, the more so for large scales, which is where inflation is expected to be the most helpful.  We had undertaken this investigation with the hope to shed some light unto the details for which this is so, and to be able to point out which of the gears operating in the mechanism are responsible for its ultimate failure.

\paragraph{The r\^ole of $E$.} First, in all cases analysed above we have observed how the different powers appearing in the expressions for the electric and magnetic energy densities result in the prevalence of the former over the latter, and its ultimate fundamental r\^ole in stopping inflation while $\delta_B$ is still small.  We can schematically formalise this finding as follows.  The electric and magnetic fields, in Fourier space, are related to the vector potential $\cA_h$ as
\be\label{EandB}
E_k^2 \simeq |\partial_\eta \cA_h|^2 \, ,\quad B_k^2 \simeq k^2 |\cA_h|^2 \, .
\ee
Normally we parametrise the vector potential as $\cA_h = y(\eta) e^{-ik\eta}$, which means that
\be\label{EkBk}
E_k^2 \simeq \left[ 1 + \left( \frac{y'}{ky} \right)^2 \right] B_k^2 \, .
\ee
Now, typically $y' \simeq y/\eta$, which translates in $y'/ky \simeq 1/k\eta$, and once plugged into the expression for $\rho_\gamma$, Eq.~(\ref{energy}), says
\be\label{energyE}
\rho_E^e \simeq \int \dd \ln k E_k^{2,e} \simeq \int \dd \ln k \left[ 1 + \left( \frac{k_\text{max}}{k} \right)^2 \right] B_k^{2,e} \, .
\ee
We can formally write $\int \dd \ln k B_k^{2,e} = B^{2,e}$.  Normally we also parametrise the $k$-dependence of $B_k$ in terms of a power law, for instance $k^n$, and, up to some order one coefficients, we obtain
\be\label{energyEfinal}
\rho_E^e &\simeq& B_\text{max}^{2,e} - B_\text{min}^{2,e} + C (B_\text{max}^{2,e} - e^{2N_t} B_\text{min}^{2,e}) \nonumber\\
&\simeq& (1 - e^{-2bN_t}) B_\text{max}^{2,e} + C(1 - e^{2(1-b)N_t})B_\text{max}^{2,e} \, ,
\ee
where the `max' and `min' subscripts as usual refer to the maximal and minimal mode to be taken into consideration and $C$ is an order 1 constant of integration; notice that at the other end of the spectrum $B_\text{min}^e = e^{-bN_t}B_\text{max}^e$.  The first term is common to both the electric and the magnetic field, whereas the second piece is peculiar to $\rho_E$ only.  If we pick a blue spectrum for the magnetic field, that is, $b>0$, we see that either the electric field is also blue (for $b>1$) and their energy densities are comparable, or, at smaller $b$ it is red, and steals most of the energy density of $\rho_\gamma$ because of the last exponential.  If the magnetic field has a red spectrum then, which is what one wishes to fit observations, the $\rho_E^e$ is always dominating by a factor of $e^{2N_t}$.  This is the source of the problem: one can not simply get away with some sizeable, red tilted, magnetic field energy density without a much stronger electric field.

The possible way to avoid this is to consider coupling constants which result in exponential amplification of the $\cA_h$, leading to $y(\eta) \simeq e^{k\eta}$ or some more complicated exponentials, in which case the precise form and coefficients appearing in the exponential will determine the winner in the electric versus magnetic competition.  For instance, a Gaussian type of evolution, $y \simeq \exp(k\eta)^2$ would reverse the situation in favour of the magnetic field.

\paragraph{Low scale inflation.} One possible place of intervention to cure these models would be the overall scale of inflation, parametrised by either $H_I$ or $a_e$. Indeed, a much lower scale of inflation would imply much less dilution until today for the magnetic power, which in turn translates into less magnetic field wanted at the end of inflation.

However, as we lower the scale of inflation, while the required magnetic strength (squared) $\delta_B^2$ decreases as $1/a_e^4$, at the same time the final energy density of inflation $\rho_\cph$ also decays.  As long as the Universe was always dominated by radiation after inflation, and the reheating process is instantaneous and has no effects on the background dynamics, then also $\rho_\cph$ follows the $1/a_e^4$ law, thereby nullifying the positive effects of the lower $H_I$.  Notice that the late-time matter dominated era does not change significantly this result, for which instead a significantly long lasting period of, say, matter dominated reheating (or any other background behaviour slower than $1/a^4$) would be needed.

One further slight improvement brought about by this possibility is that the required minimal number of e-foldings lowers, since there is less of a stretch between the highest (now lower) mode and today's Hubble scale.  As it is clear from the figures of the previous sections however, such small, in terms of the actual values involved, adjustments do not appear feasible in restoring the effectiveness of these models -- we are attempting at bridging many orders of magnitude gap in $\delta_B^0$ by means of order one or ten corrections. This avenue hence does not seem to be harbinger of any significant progress.

\subsection{A spiky counter-example}\label{spiky}

In light of the insight thus far developed, one is led to believe that if we were able to cut the spectrum of amplified modes down to a very narrow band centred around the modes of most interest, then we could be able to work our way around backreactions; moreover, one would like to keep the electric field under control while still allowing for a large magnetic field.  These two characteristics are naturally obtained from models where the growth of the $\cA_h$ is resonant, that is, $\cA_h \simeq e^{\mu\omega\Delta\eta}$, where the time span $\Delta\eta$ depends on the details of the coupling and $\omega$ is the frequency of the resonance.  Such models can be constructed from axial couplings quite easily~\cite{Byrnes:2011aa}, and have the further attractive feature that the coupling constant never needs to be large for the resonances to be efficient~\footnote{Interesting bounds on this type of coupling come from non-gaussianity, see, e.g.,~\cite{Barnaby:2010vf}.}.

Despite this being the most favourable situation one could possibly imagine, the fact that the resonance is confined to be operative within the Hubble scale is an inherent issue which ultimately will determine the demise of this kind of models.  To prove this, we will assume that
\begin{enumerate}
	\item the amplification mechanism is operative only inside the horizon, and ceases, with consequent freezing of the vector potential, once a mode becomes superhorizon;
	\item the only effect of an eventual reheating stage following the de Sitter epoch is only to alter the overall decay of the total inflaton energy density, and has no impact on the electromagnetic field;
	\item backreactions are of the background type (they impact the Friedmann equations) -- this unlaces the following estimates from a particular coupling term, and is a conservative hypothesis;
	\item the spectrum is a spike at a given wavelength; notice that this represents the best case scenario, since any different spectrum would necessarily have a higher energy density associated with it (at the same power for the pivotal wavelength) and would make the backreactions problem worse.
\end{enumerate}

In the case of a delta function type of spectrum, the magnetic energy density $\rho_B$ and the power spectrum $\delta_B^2$ coincide up to a factor of 4, and we can use either one of them in our estimates.  Now, assume that we are interested in a given final value for $\delta_B^{2,e}$ stored primarily at a scale which, for illustrative purposes, we take to be $k = \hat k = 1/\text{Mpc}$.  If nothing happens to the magnetic field from the end of inflation until today, we can estimate its value as
\be\label{b0evol}
\rho_B^e = \rho_B^0 \left( \frac{a_e}{a_0} \right)^{-4} \, .
\ee
The value of $\rho_B^0$ is what we would like to obtain from the interaction with the rolling inflaton, and can be chosen to be the observed value at that particular lengthscale, in this case 1 nGauss (squared), or a smaller value which would trigger, e.g., a dynamo amplification, say $10^{-20}$ Gauss.  If this field is frozen from the moment it is stretched beyond the Hubble horizon (as per assumption), this field will continue its evolution backwards into the inflationary epoch as
\be\label{bNevol}
\rho_B(N) = \rho_B^e e^{4N} \, .
\ee

The scale of inflation, determined by $a_e$, can be estimated using entropy conservation; we can allow for a period of matter domination behaviour after the end of inflation (reheating) when the scalar field is oscillating, and assuming all energy density of the Universe originates from the decay of the field itself, one can estimate
\be\label{infEND}
\rho_\cph^e \simeq \left( \frac{a_e}{a_r} \right)^{-3} \left( \frac{a_r}{a_0} \right)^{-4} \rho_\gamma^0 \, ,
\ee
where $a_r$ refers to the scale at which radiation domination kicks in.  If we then look at the end of the de Sitter stage, we find a ratio
\be\label{ratioEND}
\left. \frac{\rho_B}{\rho_\cph}\right|_e = x_e \simeq \frac{a_r}{a_e} \frac{\rho_B^0}{\rho_\gamma^0} = \frac{T_e}{T_r} \frac{\rho_B^0}{\rho_\gamma^0} \, .
\ee
This ratio, going back into the actual accelerated expansion epoch, will, within our approximations, grow as the fourth power of $\eta = \eta_e e^N$.  The final ingredient necessary is the number of e-folds at horizon exit, for a given mode $k$, and is given by
\be\label{Nexit}
N_k = \ln\left( \frac{\cH_e}{k} \right) \, .
\ee

There are two alternative ways to demonstrate that in this situation, the electromagnetic field (which is essentially the magnetic field because $|\partial_\eta \cA_h|^2 \simeq \mu^2 \omega^2 |\cA_h|^2$, $\mu\ll1$) is doomed to backreact before it can reach the desired values.  One can decide on a particular value $\rho_B^0$, follow it up until horizon crossing given by~(\ref{Nexit}) and compare it to $\rho_\cph$ at the same time: if this ratio is larger than one than it means that the electromagnetic field has to be larger than what is allowed to be at horizon exit, up until when its evolution is certain.  Similarly, one can compute the number of e-folds at which the ratio $x = x(N) = 1$ and if this $N$ turns out to be smaller than $N_k$ then the system backreacts.

The second possibility is to choose the most optimistic option $\rho_B(N_k) = \rho_\cph(N_k)$, and follow the magnetic energy density down to today.  This leads to
\be\label{rhoB0best}
\rho_B^0 \simeq \frac{3\left( 1+4N_t/\alpha \right)^{\alpha/2+1}}{50\pi^2\delta_H^2} k^4 \, ,
\ee
where we have taken $\rho_\cph(N) = \rho_\cph^e$ for simplicity.  This expression~(\ref{rhoB0best}) relates the largest possible magnetic field at a given scale $k$ today with the duration of inflation, and, for large modes especially, clearly shows how any reasonable duration for inflation automatically leaves us with only crumbs of magnetic energy today: 66 e-foldings of chaotic ($\alpha=2$) inflation can give birth to a disheartening $10{-51}$ Gauss at 1 Mpc today.

\section{Conclusions}\label{wrapup}

To conclude, we have analysed the impact of the backreactions during inflation, when the inflaton field itself is coupled to the electromagnetic vector potential.  Thanks to a choice of physically motivated parametrisations, we were able to compare the allowed regions in the parameters spaces which lead to a controlled background, and at the same time compute the strength of the magnetic field today at any given interesting wavelength, without having to rely on a specific type of coupling, and to a certain extent, independently on the details of inflation.

We primarily focussed on large scales, as they are the ones which appear to be harder to be generated via late-time mechanisms such as phase transitions or astrophysical processes.  The results are somewhat discouraging, as it was essentially impossible to find a parametrisation and a choice of values for the parameters which allowed any significant production of magnetic fields.  We have elaborated on four possibilities: simple growth with different strength inside and outside the Hubble scale; amplification within a narrow window around horizon crossing; power-law mode-dependent growth; spectral cuts, both infrared and ultraviolet.  In all these cases the strength of the magnetic fields today could never exceed $10^{-30}$ Gauss on Mpc scales.

We have found that the chief reason behind this result is to be found in the presence of the \emph{electric} field alongside the magnetic counterpart; the different power of momentum with which the former appears in the energy density makes its spectrum always redder than that of the magnetic field, for the same values of the parameters.  Therefore, unless a very specific mechanism is able to differentiate between the two components by boosting the magnetic field while leaving the electric one alone, large scale power will always need to be suppressed.  Also, the fact that normally all modes are amplified democratically, with no distinction among different energy scales, in conjunction with the width of the spectrum accessible by inflation, makes for a challenging selection of large scales to be appointed a significantly large power.  While some of these considerations were available in the literature, scattered among a handful of papers, we have here tackled this issue in an organic way.

This constitutes only a first step towards a more complete understanding of ``what goes wrong'' when we attempt to generate magnetic fields during inflation: being able to point out where the backreactions arise from in a model-independent way is a necessary starting point in building a successful model where this can be accomplished.  Of course we have not analysed what actually happens when the electromagnetic field energy density approaches that of the background, and the ultimate fate of inflation depends on the details of the coupling through the coupled equations of motion.  Furthermore, we have not dug in the reheating stage following inflation, where the magnetic power spectrum can change shape and intensity quite dramatically, as we leave this much more involved issue for future investigation.

However, although we could conceive an anaesthetic combination of the elements examined in this work which might succeed in the very end, our order of magnitude analysis showed how, on fairly general and solid grounds, if inflation were to survive despite the menacing electromagnetic field, the resulting magnetic half would be completely negligible, the more so on large scales, thereby precluding the identification of such fields with the observed large scale magnetism.

\section*{Acknowledgements}

I wish to thank L~Hollenstein and C~Byrnes for valuable discussions and for collaborating at the early stages of this work.

\bibliography{NoGo}
\bibliographystyle{JHEP}

\end{document}